\documentclass[10pt]{article}
\usepackage{opex3}
\usepackage{amssymb, amsmath, bbm}
\newcommand{\unit}[1]{\ensuremath{\, \mathrm{#1}}}
\newcommand{\bra}[1]{\langle #1|}
\newcommand{\ket}[1]{|#1\rangle}
\newcommand{\braket}[1]{\langle #1 \rangle}

\begin{document}

\title{Witnessing effective entanglement\\ over a 2$\,$km fiber channel}

\author{Christoffer Wittmann$^{1,2.\dagger}$, Josef F\"urst$^{1,2,\dagger}$, Carlos Wiechers$^{1,2,3}$, Dominique Elser$^{1,2}$, Hauke H\"aseler$^{2,4}$,\\ Norbert L\"utkenhaus$^{1,2,4}$, and Gerd Leuchs$^{1,2}$}

\address{$^{1}$Max Planck Institute for the Science of Light,\\ G\"unther-Scharowsky-Str. 1 / Bau 24, 91058 Erlangen, Germany\\
$^{2}$Institute of Optics, Information and Photonics,\\ University Erlangen-Nuremberg, Staudtstra\ss e 7/B2, 91058 Erlangen, Germany\\
$^3$ Departamento de F\'{\i}sica, Universidad de Guanajuato, 
Lomas del Bosque 103,\\ Fraccionamiento Lomas del Campestre, 
37150, Le\'{o}n, Guanajuato, M\'{e}xico\\
$^{4}$Institute for Quantum Computing, University of Waterloo,\\
200 University Avenue W., Waterloo, Ontario, Canada N2L 3G1\\
$\dagger$ Authors contributed equally to this work.}

\email{Christoffer.Wittmann@mpl.mpg.de} 



\begin{abstract}
We present a fiber-based continuous-variable quantum key distribution system. In the scheme, a quantum signal of two non-orthogonal weak optical coherent states is sent through a fiber-based quantum channel. The receiver simultaneously measures conjugate quadratures of the light using two homodyne detectors. From the measured Q-function of the transmitted signal, we estimate the attenuation and the excess noise caused by the channel. The estimated excess noise originating from the channel and the channel attenuation including the quantum efficiency of the detection setup is investigated with respect to the detection of effective entanglement. The local oscillator is considered in the verification. We witness effective entanglement with a channel length of up to $2 \unit{km}$.
\end{abstract}




\section{Introduction}

``Kerckhoffs' principle'' \cite{kerckhoffs_la_1883} and Shannon's assumption ``The enemy knows the system'' \cite{shannon_communication_1949} established the basis for modern cryptography and enhanced secure communication between two parties.
With the proposal of the ``one time pad'' \cite{vernam_cipher_1926}, the security aspect shifted from secure communication to secure key distribution between these two parties. Quantum key distribution (QKD), first proposed 1984 \cite{bennett_quantum_1984}, offers a way to exchange a secret key using the quantum mechanical properties of light as the carrier of information. The security is thereby based on fundamental physical concepts. 

First proof of principle experiments for QKD \cite{bennett_experimental_1992, ekert_practical_1992, muller_experimental_1993} were followed by practical implementations over long and extremely long distances \cite{stucki_quantum_2002, rosenberg_long-distance_2007, ursin_entanglement-based_2007, schmitt-manderbach_experimental_2007}. In parallel to these discrete-variable QKD systems, continuous-variable QKD using homodyne detection was proposed~\cite{ralph_continuous_1999}. It was shown, that the limit of 3\,dB channel attenuation can be overcome by the concept of postselection~\cite{silberhorn_continuous_2002} and the postprocessing method reverse reconciliation~\cite{grosshans_quantum_2003}. Continuous-variable QKD has  been tested on quantum channels of up to 25\.km length~\cite{lodewyck_quantum_2007, qi_experimental_2007} using homodyne detection and basis switching. Besides using a single homodyne detection, also simultaneous detection of both conjugate quadratures of the signal was demonstrated~\cite{lorenz_continuous-variable_2004, lance_no-switching_2005}. Homodyne detection of conjugate quadratures referred to as heterodyne detection is particularly interesting for three reasons:
\begin{itemize}
	\item Random numbers are not needed in the receiver's setup. 
	\item The Trojan-horse attack~\cite{gisin_trojan-horse_2006}, where Eve gains information by reading the basis choice in Bob's setup, is not possible. 
	\item The heterodyne detection strategy achieves higher secure bit rates than schemes with homodyne detection in some QKD protocols~\cite{pirandola_continuous-variable_2008,weedbrook_coherent-state_2006,lodewyck_tight_2007}.
\end{itemize}
Experiments using heterodyne detection on a long fiber channel, however, had not yet been demonstrated.

In this paper, we present a fiber-based QKD-system using a double-homodyne detection setup. As in previous experiments \cite{lorenz_continuous-variable_2004, elser_feasibility_2009}, we use binary encoded continuous-variable quantum states consisting of a signal mode and a local oscillator mode. We adapt our previous polarization-based experiments to a fiber channel. To this end, our two mode states are sent through the quantum channel using a combination of time~\cite{lodewyck_quantum_2007} and polarization multiplexing~\cite{qi_experimental_2007} of signal and local oscillator (LO). Therefore, detrimental effects from photon-phonon interactions (GAWBS) are avoided~\cite{elser_guided_2007}. By describing the sent quantum information in the Stokes space~\cite{korolkova_polarization_2002}, it is possible to verify effective entanglement in the measurement data according to the method in~\cite{hseler_testing_2008}. In contrast to all previous experiments, not only the signal but also the strong reference beam (LO) is considered in this security analysis. We show how to measure conjugate Stokes parameters with a freely drifting interferometric phase. The phase drift is monitored by additional classical calibration pulses on the quantum channel. Subsequently, Bob remaps his measurement data, similar to~\cite{qi_experimental_2007}. Finally, monitoring the intensity of the local oscillator at the detection stage allows us to unambiguously demonstrate the generation of quantum-correlated data.

The paper is organized as follows. In Section~\ref{protocol}, we introduce our protocol and the Stokes formalism. In Section~\ref{experiment}, we describe the setup focusing on the implementation of the detection part. In Section~\ref{verification}, noise characteristics of the system and the transmitted signal states are presented.  Finally, the entanglement criterion is applied.

\section{The protocol}
\label{protocol}

\begin{figure}
\centering\includegraphics[width=0.9\textwidth]{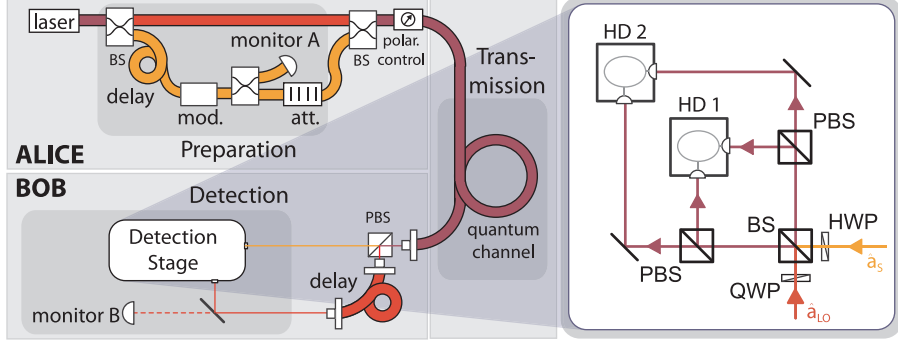}
\caption{\label{fig:setup}(left) Schematics of the QKD setup, (right) the detection part is a random phase heterodyne detection using the polarization degree of freedom. Our scheme is similar to a technique called linear optical sampling~\cite{dorrer_linear_2003}. In contrast to their system, we have a shot noise limited detection and a SNR improvement of about 40\,dB. }
\end{figure}

Our protocol is a continuous-variable adaptation of the two-state protocol proposed by Bennett~\cite{bennett_quantum_1992-1}. Alice wishes to share a random bit string with Bob and starts by encoding binary information in the phase of coherent states. These two signal states $\ket{\alpha}$ and $\ket{-\alpha}$ are transmitted through a quantum channel and evolve to the output states $\rho_0$ and $\rho_1$, which are detected by Bob in a heterodyne setup. Bob records the expectation values and variances of two orthogonal quadratures, which allows him to reconstruct a bit value from each signal and to estimate the information potentially extracted by an eavesdropper. A detailed account of the classical data processing necessary to distill a secret key from the raw measurement outcomes is presented in Ref.~\cite{zhao_asymptotic_2009}.

This two-state protocol is attractive from an experimental point of view, since the signal preparation is the simplest possible. Theoretically, lower bounds on the achievable secret key rate have been evaluated in Ref.~\cite{zhao_asymptotic_2009}. Unfortunately, those results are not applicable to practical scenarios, since the key rate drops too quickly with increasing excess noise. The theory needs further development to make the protocol more robust against noise.

Here, we show that our experiment can establish quantum correlations in the classical data held by Alice and Bob, which is a necessary precondition for the distillation of a secret key~\cite{curty_entanglement_2004}. The existence of quantum correlations is proven through the verification of so-called \emph{effective entanglement} \cite{bennett_quantum_1992}, a theoretically constructed, alternative description of the signal source. In this description,  Alice prepares bipartite states
\begin{equation}\label{eq:entbased}
	\ket{\Psi}_\mathrm{AB} = \frac{1}{\sqrt{2}} ( \ket{0}_\mathrm{A} \ket{\alpha}_\mathrm{B} + \ket{1}_\mathrm{A} \ket{-\alpha}_\mathrm{B} ).
\end{equation}
Her subsequent orthogonal projective measurements $\{ \ket{0}_A \bra{0}, \ket{1}_A \bra{1} \}$ effectively prepare the signal states $\ket{\pm\alpha}_B$, with equal probability.  The action of the quantum channel on the state $\ket{\Psi}_{AB}$ leads to an effective bipartite quantum state $\rho_{AB}$ shared by Alice and Bob. It is their task to prove that the classical measurement data must come exclusively from entangled $\rho_{AB}$, i.e., their data is quantum correlated.

Due to the discrete-continuous structure of the setup, we use the Expectation Value Matrix method~\cite{hseler_testing_2008, rigas_entanglement_2006} to verify the existence of quantum correlations. In this method, all measured expectation values are stored in a matrix, and the separability of this matrix is linked to the separability of the possible underlying quantum states. For a detailed description, the reader is referred to Ref.~\cite{hseler_testing_2008}. Reference~\cite{hseler_testing_2008} also reveals a vulnerability of the protocol arising from an attack on the local oscillator mode. In a homodyne setup, the assumption that the local oscillator is a strong coherent state allows us to regard the measurement as the single-mode detection of the signal's quadrature~\cite{leonhardt_measuringquantum_1997}. Precisely this assumption allows Eve to perform an intercept-resend attack which modifies the intensities of signal and local oscillator modes. Such an attack will not show in Bob's quadrature detection, but could lead to data which is, at best, classically correlated and therefore not suitable for key generation.

\begin{figure}
	\centering
		\includegraphics[width=0.7\textwidth]{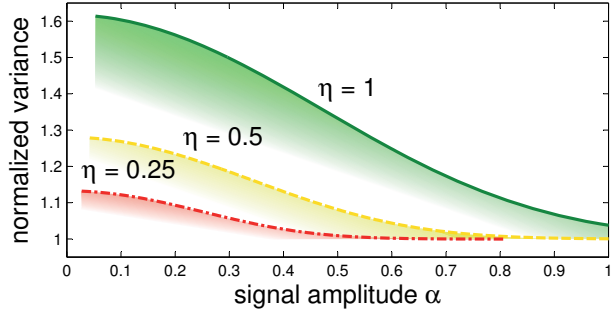}
	\caption{\label{fig:theory}Curves show lower bounds for quantum correlations. The variance of the Stokes operators is normalized by the local oscillator intensity resulting in the normalized variance $\mathrm{Var}(\hat S_{2,3})/\braket{\hat n_\mathrm{LO}}$. The different curves correspond to different channel transmissions $\eta$. Effective entanglement is verified in the shaded areas.}
	
\end{figure}

To take this into account, we consider the binary signals as true two-mode states $\ket{\pm \alpha} \otimes \ket{\alpha_{LO}}$ and we regard the homodyne detection as a measurement of the quantum Stokes operators
\begin{eqnarray}
\label{Stokesoperators1}
	\hat S_1&=&\hat a^\dagger_{LO} \hat a_{LO} - \hat a^\dagger_S \hat a_S = \hat{n}_{LO} - \hat{n}_s \\\nonumber
	\hat S_2&=&\hat a^\dagger_{LO} \hat a_S + \hat a^\dagger_S \hat a_{LO}\\\nonumber
	\hat S_3&=&i(\hat a^\dagger_{S} \hat a_{LO} - \hat a^\dagger_{LO} \hat a_S).
\end{eqnarray}
The protocol gives access to the expectation values and variances of two of the Stokes operators, say $\hat{S}_2$ and $\hat{S}_3$. Then, a suitable Expectation Value Matrix is constructed as follows~\cite{hseler_testing_2008}:
\begin{equation}
 \chi(\rho_{AB}) = 
  \begin{bmatrix}
    \Big\langle |0\rangle\langle 0|\otimes B \Big\rangle_{\rho_{AB}} & 
   \Big\langle |0\rangle\langle 1|\otimes B \Big\rangle_{\rho_{AB}} \\
    \Big\langle |1\rangle\langle 0|\otimes B \Big\rangle_{\rho_{AB}} & 
 \Big\langle |1\rangle\langle 1|\otimes B \Big\rangle_{\rho_{AB}}
  \end{bmatrix}
\end{equation}
with
\begin{equation}
B=\begin{bmatrix}
\hat{1}_B & \hat{S}_2 & \hat{S}_3\\
\hat{S}_2 & \hat{S}_2^2 & \hat{S}_2\hat{S}_3\\
\hat{S}_3 & \hat{S}_3\hat{S}_2 & \hat{S}_3^2
\end{bmatrix}.
\end{equation}
Separability of the underlying state is then tested through the condition
\begin{equation}
	\chi(\rho_{AB})^{T_A} \ge 0,
\end{equation}
and violation of this certifies effective entanglement. The remaining Stokes operator $\hat{S}_1$ enters the Expectation Value Matrix through the symmetrizing step $\hat{S}_2\hat{S}_3 = (\hat{S}_2\hat{S}_3 + \hat{S}_3\hat{S}_2 )/2 + (\hat{S}_2\hat{S}_3 - \hat{S}_3\hat{S}_2 )/2 = \{ \hat{S}_2, \hat{S}_3 \}/2 + i \hat{S}_1$, which is instrumental to the verification process. However, due to practical reasons, the expectation value $\braket{\hat{S}_1}$ is not directly measured in our experiment, and instead, we measure the photon number in the local oscillator mode (monitor B in Fig.~\ref{fig:setup}(left)). We now show how this detour can provide a lower bound on $\braket{\hat{S}_1}$:

Direct calculation leads to the relation
\begin{equation}\label{eq:st1}
	\hat{S}_2^2 + \hat{S}_3^2 = 2 (\hat{S}_0 + 2 \hat{n}_{LO} \hat{n}_s),
\end{equation}
where $\hat{S}_0$ is defined as the total intensity $\hat{n}_s + \hat{n}_{LO}$. Equation (\ref{eq:st1}) contains two quantities which are not directly accessible in our experiment, namely $\hat{S}_0$ and $\hat{n}_s$. The latter is easily replaced by the relation
\begin{equation}
	\hat{n}_s = \frac{1}{2} ( \hat{S}_0 - \hat{S}_1 ),
\end{equation}
leading to
\begin{equation}
	\hat{S}_1 = \hat{S}_0 (1 + \frac{1}{\hat{n}_{LO}} ) - \frac{\hat{S}_2^2 + \hat{S}_3^2}{2 \hat{n}_{LO}}.
\end{equation}
Finally, we use the relation $\braket{\hat{S}_0} \ge \braket{\hat{n}_{LO}}$ to arrive at the desired bound
\begin{equation}
\label{S1est}
	\braket{\hat{S}_1} \ge 1 + \braket{\hat{n}_{LO}} - \frac{ \braket{\hat{S}_2^2 + \hat{S}_3^2} }{ 2 \braket{\hat{n}_{LO}} }.
\end{equation}
At this point, we can directly apply the Expectation Value Matrix method to verify the existence of quantum correlations, using the approximation (\ref{S1est}) instead of the actual value of $\braket{\hat{S}_1}$. Figure \ref{fig:theory} shows the noise levels for different choices of input intensities, under the assumption that the noise is equal for both quadratures and both signal states. The variances are normalized by $\braket{\hat{n}_{LO}}$ to make the results independent of the local oscillator intensity. We observe that more noise can be tolerated when smaller signal amplitudes are chosen. Channel losses decrease the tolerable excess noise for quantum-correlated data.

\section{Experimental Setup}
\label{experiment}

In this section, we first describe Alice's and Bob's optical hardware. We then separately discuss the detection scheme and finally explain the control software.

\subsection{Optical setup}

The optical setup of our QKD-system is shown in Fig.~\ref{fig:setup}. The Alice module consists of a diode laser (\textit{SLT5411} from \textit{Sumitomo Electronic Industries} as used in~\cite{legr_implementation_2006}) pulsed by a self-made pulsed current supply. The laser pulses are approximately 100\,ns long and have a wavelength of 1549.3\,nm. The line width is 6.6\,GHz at -10\,dB of optical power and the coherence time is estimated with self-homodyning to be 0.2\,ns. The laser pulses are split asymmetrically in the LO and the signal arm. The smaller fraction is used for the signal preparation. The signal arm consists of a delay fiber, a Mach-Zehnder modulator for amplitude modulation, a monitor detector, and an optical attenuator. This results in a shot noise limited weak signal at the single photon level. In the second fiber beamsplitter (BS) the LO pulses and the signal pulses are spatially combined with a 500\,ns time shift. Additionally, the polarization of the pulses is chosen orthogonal. The last component in the Alice setup is a computer controlled polarization controller pre-compensating the slow polarization drift in the fiber channel.

In the Bob module the signal is demultiplexed with a polarizing beamsplitter (PBS). We control the second port of the PBS with a physical block to ensure the second input mode is in a vacuum state. The LO passes a delay fiber, while the signal is directly sent to the free-space detection setup. A monitor diode behind a highly reflective mirror measures the LO energy pulse by pulse. The polarization control is set such that the power on the monitor diode is maximized. Finally the signal is measured with our detection scheme shown in (Fig.~\ref{fig:setup}~(right)).

\subsection{Detection System}
\label{Detection}

The detection system in Fig.~\ref{fig:setup}~(right) consists of two homodyne detectors detecting two conjugate quadratures of the signal mode simultaneously. This was first investigated in \cite{shapiro_phase_1984, stenholm_simultaneous_1992, leonhardt_realistic_1993}.
Our scheme consists of free-space optics and therefore allows for easy and lossless manipulations of the polarization. The LO's polarization is then chosen to be circular. The signal's polarization is tilted by 45$^\circ$ with respect to H and V polarization. Both beams interfere on a polarization independent 50:50 beam splitter with acute angle of incidence.

For the H-polarized component, we label the relative phase between the signal and the bright
LO $\phi_H=\phi_I$ , where $\phi_I$ originates from interferometer drifts in sender and receiver modules. The phase shift of $\pi/2$ between two orthogonally polarized LO components results in
a relative phase for the V-components of $\phi_V=\phi_I+\pi/2$. Both beams propagate to a PBS, which separates the H- and V-polarization. The reflected V-components impinge on the first homodyne detector (HD1), while the transmitted H-components impinge on the second homodyne detector (HD2). The difference of the photo currents is recorded.

Commonly, homodyne detectors are treated as quadrature detectors. We calculate the photon number difference $\hat n^{\mathrm{HD1}}_{-}$ detected by HD1 using the linearized field operators $\hat a_S = \alpha \cdot \hat 1 +\delta \hat a_S$ and $\hat a_{LO} =  \alpha_{LO} \cdot \hat 1 + \delta \hat a_{LO}$. We assume that $\alpha_{LO}$ and $\alpha$ are real and $\alpha_{LO} \gg \alpha$. Additionally, the modes $S_\bot$ and $LO_\bot$,  the modes orthogonally polarized to signal and LO, are in a vacuum state, i.e. $\hat a_{S_\bot} = \delta \hat a_{S_\bot}$ and $\hat a_{LO_\bot} = \delta \hat a_{LO_\bot}$.  The last assumption is justified, since the other input port of the PBS is under Bob's control. The detected signal is found after a straightforward calculation to be
\begin{eqnarray}
\label{HD1}
\hat n^\mathrm{HD1}_{-}&=&\frac{1}{2}\left((\hat a_{LO}e^{i\phi_I}+\hat a_{LO_\bot}e^{i\phi_I})^\dagger (\hat a_S+\hat a_{S_\bot})\right. \\\nonumber
&& \left. + (\hat a_S+\hat a_{S_\bot})^\dagger (\hat a_{LO}e^{i\phi_I}+\hat a_{LO_\bot}e^{i\phi_I}) \right) \\\nonumber 
&=& \alpha_{LO}\left(\cos (\phi_{I}) \cdot \alpha+\delta \hat X_{S,\phi_{I}} +\delta \hat X_{S_\bot}\right),
\end{eqnarray}
where $\hat X_{M,\phi}=\frac{1}{2}(\hat a^\dagger_M e^{i\phi}+ \hat a_M e^{-i\phi})$ is a quadrature operator for mode $M$ with a phase $\phi$.
This quadrature measurement is derived analogously for HD2. It reads
\begin{eqnarray}
\label{HD2}
\hat n^\mathrm{HD2}_{-}&=&\frac{1}{2}\left((\hat a_{LO}e^{i\phi_I+\pi/2}-\hat a_{LO_\bot}e^{i\phi_I+\pi/2})^\dagger (\hat a_S-\hat a_{S_\bot}) \right. \\\nonumber
&& \left. + (\hat a_S-\hat a_{S_\bot})^\dagger (\hat a_{LO}e^{i\phi_I+\pi/2}-\hat a_{LO_\bot}e^{i\phi_I+\pi/2}) \right) \\\nonumber 
&=& \alpha_{LO}\left(\cos (\phi_{I}+\pi/2) \cdot \alpha+\delta \hat X_{S,\phi_{I}+\pi/2} +\delta \hat X_{S_\bot} \right).
\end{eqnarray}
The vacuum input $S_\bot$ at the PBS adds additional 3\,dB noise to the variance of the measurement. Therefore our setup is equivalent to a standard heterodyne detection. We do not stabilize the interferometric phase $\phi_I$ but let it drift freely. However, we have an inherent stabilization of the relative phase of $\phi_{H}-\phi_{V}=\pi/2$.  Therefore two random but conjugate quadratures are measured with HD1 and HD2.

\begin{figure}
\begin{tabular}{l}
\centerline{\includegraphics[width=5.2cm]{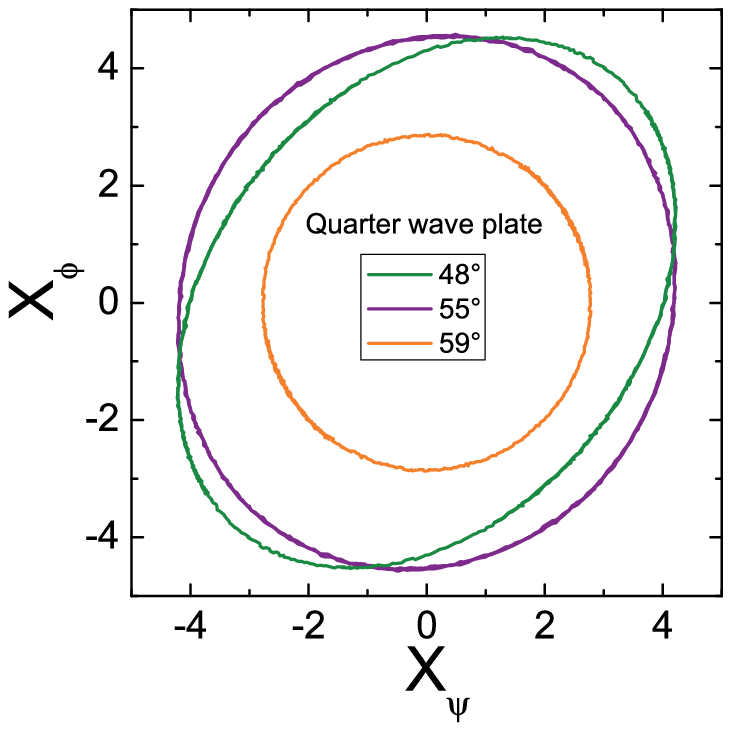}\hspace{0.3cm}\includegraphics[width=7.6cm]{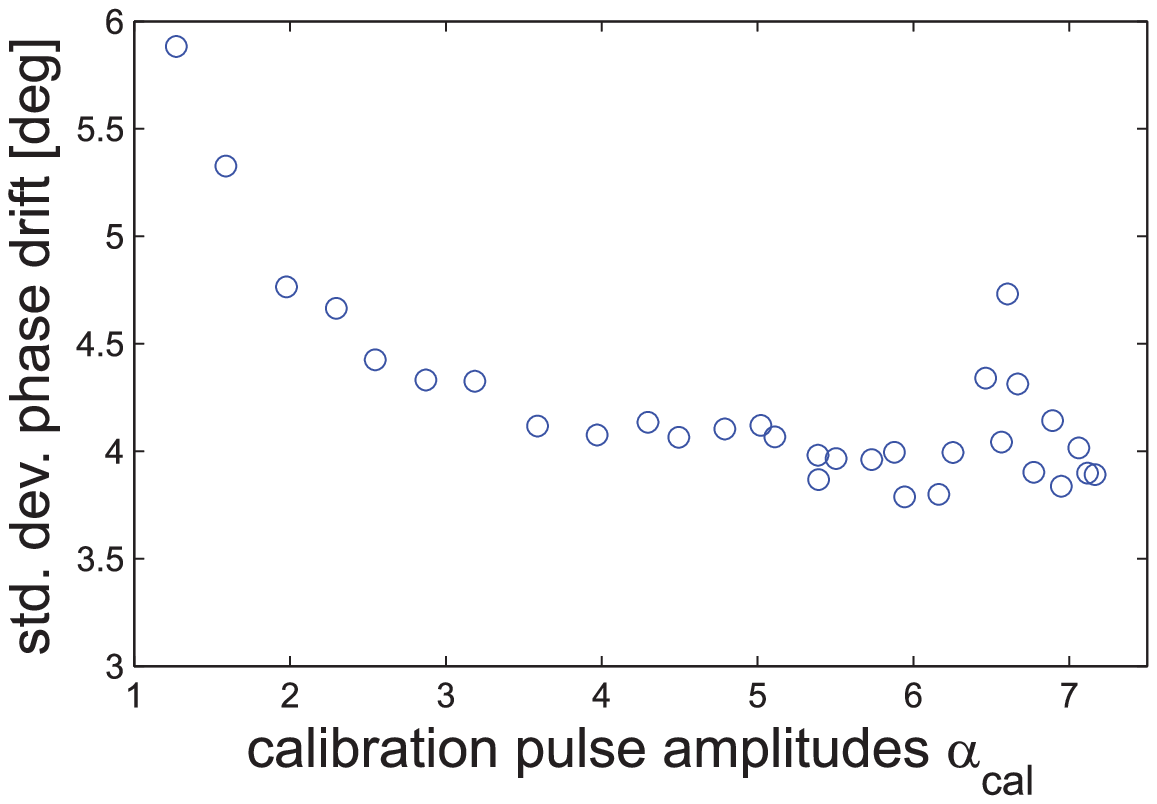}}\\[-5cm]
\hspace{0.5cm}(a)\hspace{4.2cm}(b)\\[4.3cm]
\end{tabular}
\caption{\label{fig:detection} (a) Lissajous figure of a phase randomized signal is measured for different settings of the QWP in the setup. If the measured quadratures $X_\psi$ and $X_\phi$ are not orthogonal, the shape of the graph will be elliptical. We demonstrate that orthogonal quadratures are measured for the correct QWP angle (orange trace). (b) The phase is estimated from blocks of calibration pulses. The figure shows the standard deviation of the phase drift between different calibration blocks.}
\end{figure}

The simultaneous measurement of two conjugate quadratures depends strongly on the proper choice of the polarization of signal and LO in the detection setup. For balancing, a linearly polarized LO is equally split to both homodyne detectors. A weak signal beam is then sent into the detection scheme. The quadrature measurements of 4000 signals are averaged and the mean values are plotted in the two-dimensional measurement space of Fig.~\ref{fig:detection}(a). Phase drifts of the interferometric phase will result in elliptical or circular graphs, depending on the angle of the quarter wave plate (QWP) in the LO path, corresponding to largely correlated measurements (elliptical) and uncorrelated measurements (circular). It is easy to show, that by turning the QWP, two quadratures with arbitrary relative angle can be measured. We desire uncorrelated measurements (orange trace in Fig.~\ref{fig:detection}(a)).

\subsection{Electronics and Automatization}
\label{Automatization}

In the following, we describe the electronic hardware, the control software and the steps in the postprocessing. The Alice electronics is essentially a 14-bit-D/A-converter to prepare the signal with a sampling rate of 20\,MS/s. One port drives the laser with a pulse rate of 1 MHz. Another port produces rectangular pulses for the amplitude modulator. A third port synchronizes Bob's experiment with an electronic clock signal at approximately 1\,kHz. The clock could be substituted for a synchronization using the LO monitor diode and calibration pulses as time stamp in future experiments.

A quaternary modulation is applied to the amplitude modulator to create the states $|\alpha\rangle$, $|{-}\alpha\rangle$, $|\alpha_{cal}\rangle$ and $|{-}\alpha_{cal}\rangle$, where  $\alpha$ is the signal amplitude and $\alpha_{cal}$ is the amplitude of brighter ``classical'' calibration pulses sent along with the signal. The pulse pattern consists of four calibration pulses followed by 28 signal pulses.

To pre-compensate the polarization drift, we inserted the polarization control in Alice's setup. An optimal separation of signal and LO is obtained by maximizing the LO power on the monitor diode. The power is maximized either manually or with Bob's PC using a simplex method~\cite{nelder_simplex_1965}. This demultiplexing method is very stable. It is furthermore lossless as opposed to a coupler with fixed splitting ratio~\cite{lodewyck_quantum_2007}.

In the following, we describe the electronic circuits. The difference signal in the homodyne detectors is amplified by charge sensitive amplifiers as in the design by \cite{hansen_ultrasensitive_2001}. The electrical pulse duration produced by the detectors is set to 400\,ns (foot), which allows for repetition rates up to 2\,MHz while maintaining linear amplification. The linearity was confirmed  for all four photodiodes independently. Due to limitations of computing power, and an electronic signal due to LO light leaking into the signal arm, we run the experiment at 1\,MHz. When varying the LO power for balanced homodyne detection of vacuum, we find a linear behavior of the signal variance versus the LO intensity. For typical LO power of $10^8 \,\textrm{photons}/\textrm{pulse}$, the electronic noise is 20 dB below the signals variance. The common mode rejection ratio is always better that 40\,dB.
The detection efficiency of the homodyne detectors was 70\%, including the quantum efficiency of the diodes (86\%), the mode matching efficiency ($95.4\%$) and the loss in optical components (10\%).

Bob's 12-bit-A/D-converter digitizes the signal and the LO monitor detectors with 16\,MS/s. We reduce the number of samples by neglecting samples in between the pulses \footnote{In a QKD system the removed samples should be checked for manipulation attacks.} and averaging 8 samples (approximately the electronic pulse length). These mean values are used to estimate the shot noise level, as shown in Eqn.~(\ref{S1est}). 
In the postprocessing, we estimate the mean value of the Stokes operators $\hat S_2$ and $\hat S_3$ for 1024 signal states. The displacement with respect to this mean value is considered as quantum signal. Long term fluctuations of the detector are thereby compensated. 

The interferometric phase $\phi_I$ is estimated with the four bright calibration pulses in each 32-pulse-frame. To measure the phase noise, we calculate the phase drift between two calibration steps. We show the standard deviation of the phase drift in Fig.~\ref{fig:detection}(b). We find that for weak calibration pulses the standard deviation depends on the calibration pulse amplitude. This stems from the limit for the phase estimation of weak coherent states~\cite{leonhardt_canonical_1995}. For stronger amplitudes, the standard deviation of the phase drift is measured to be 4 degree for calibration times of $32\,\mu$s.  With the estimated phase, we remap the coordinate system of the measured frame to the phase space of Alice's signal states. Subsequently, the data can be analyzed as described in the next section. The computational power needed for the complete postprocessing is high, but at 1MHz repetition rate the system runs continuously in realtime, as required for practical use in a QKD system.

\section{Verification of Effective Entanglement}
\label{verification}

\begin{figure}
\begin{tabular}{l}
\centerline{\includegraphics[width=6.2cm]{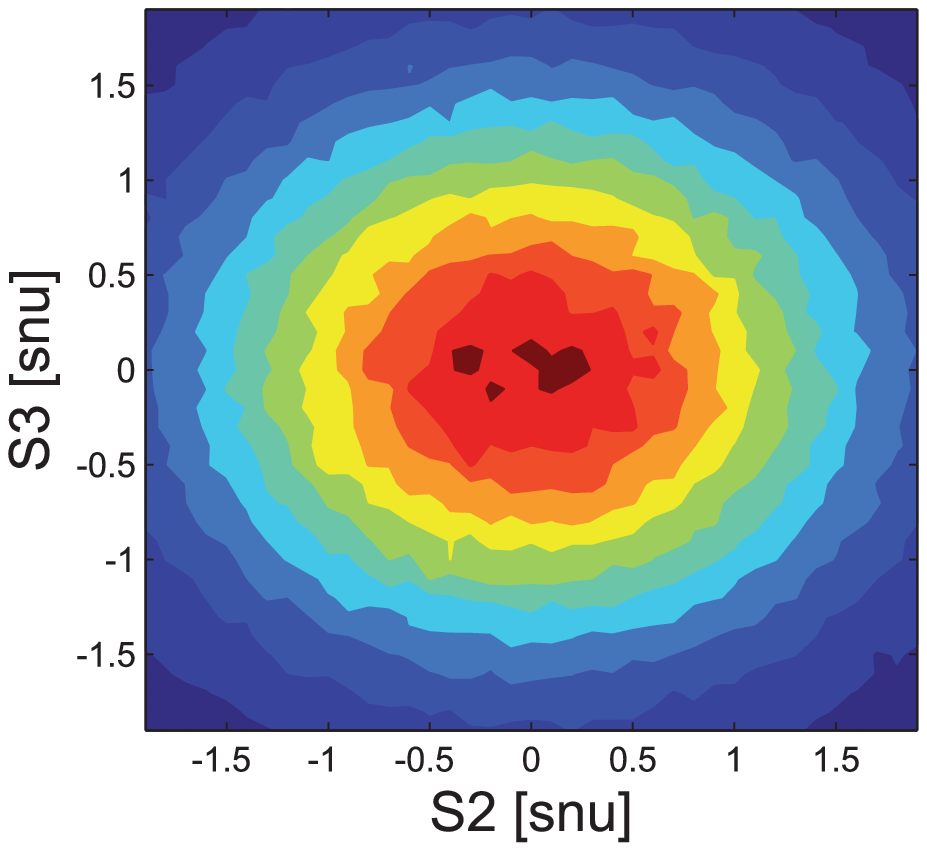}\hspace{0.3cm}\includegraphics[width=6.2cm]{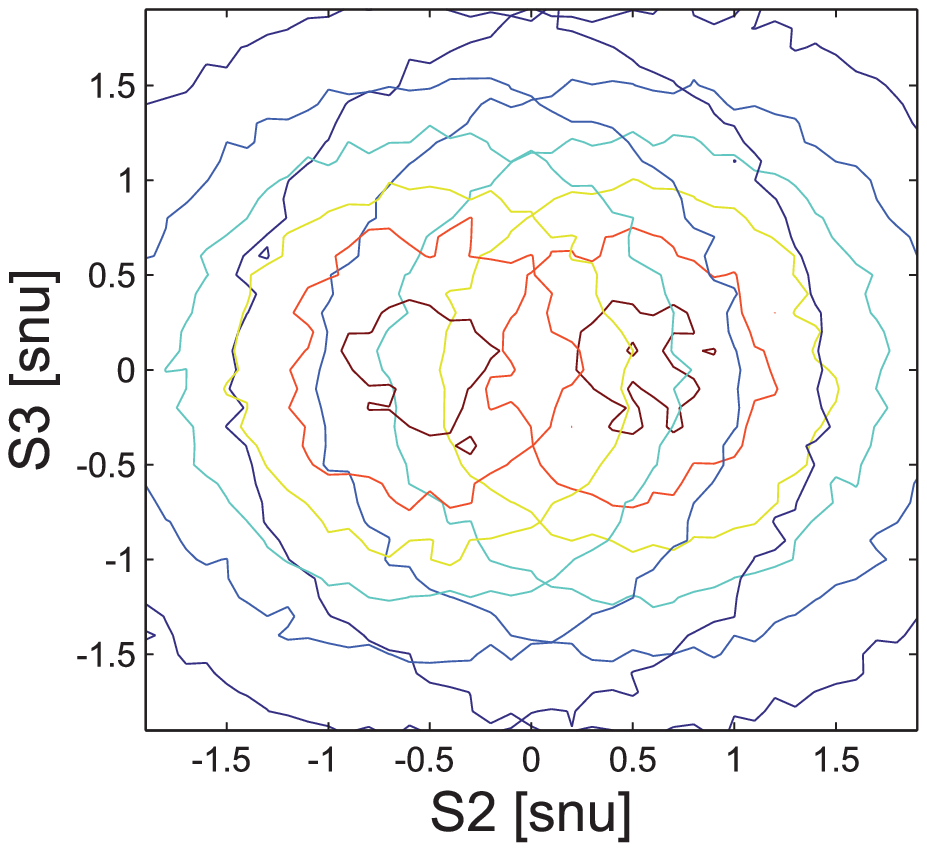}}\\ \centerline{\includegraphics[width=7.0cm]{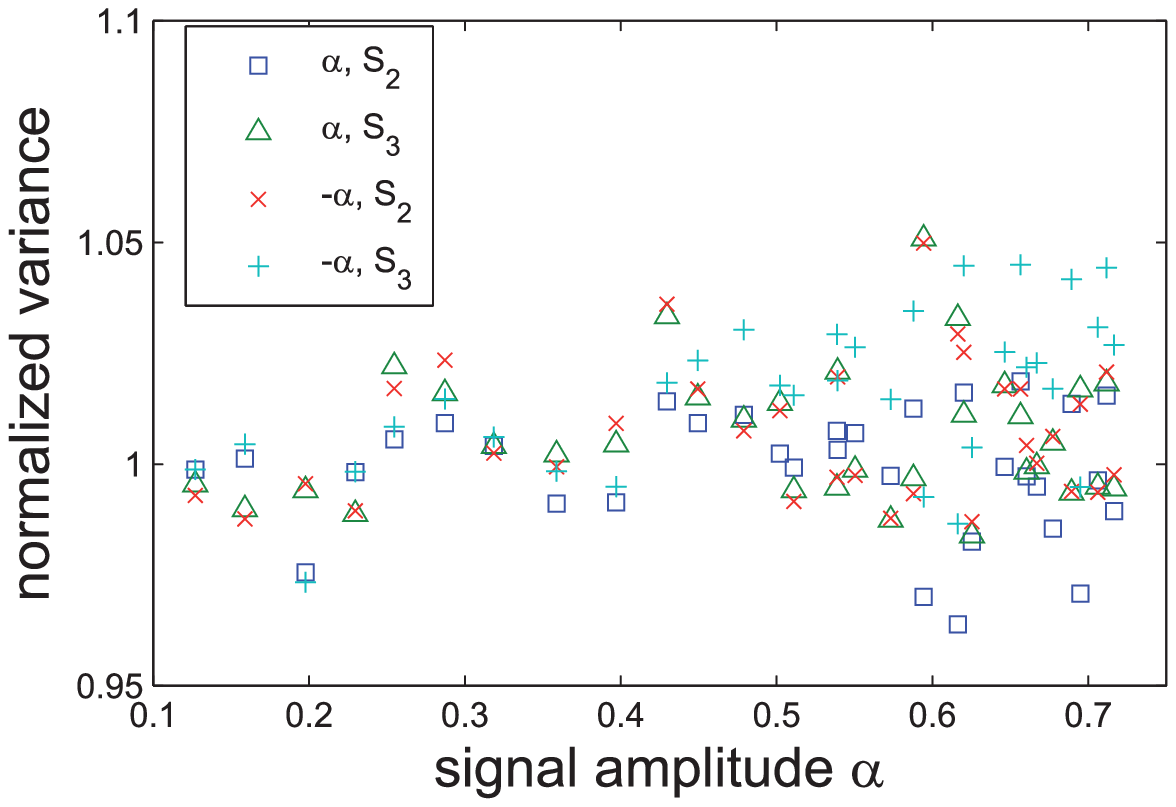}\hspace{-0.4cm}\includegraphics[width=7.0cm]{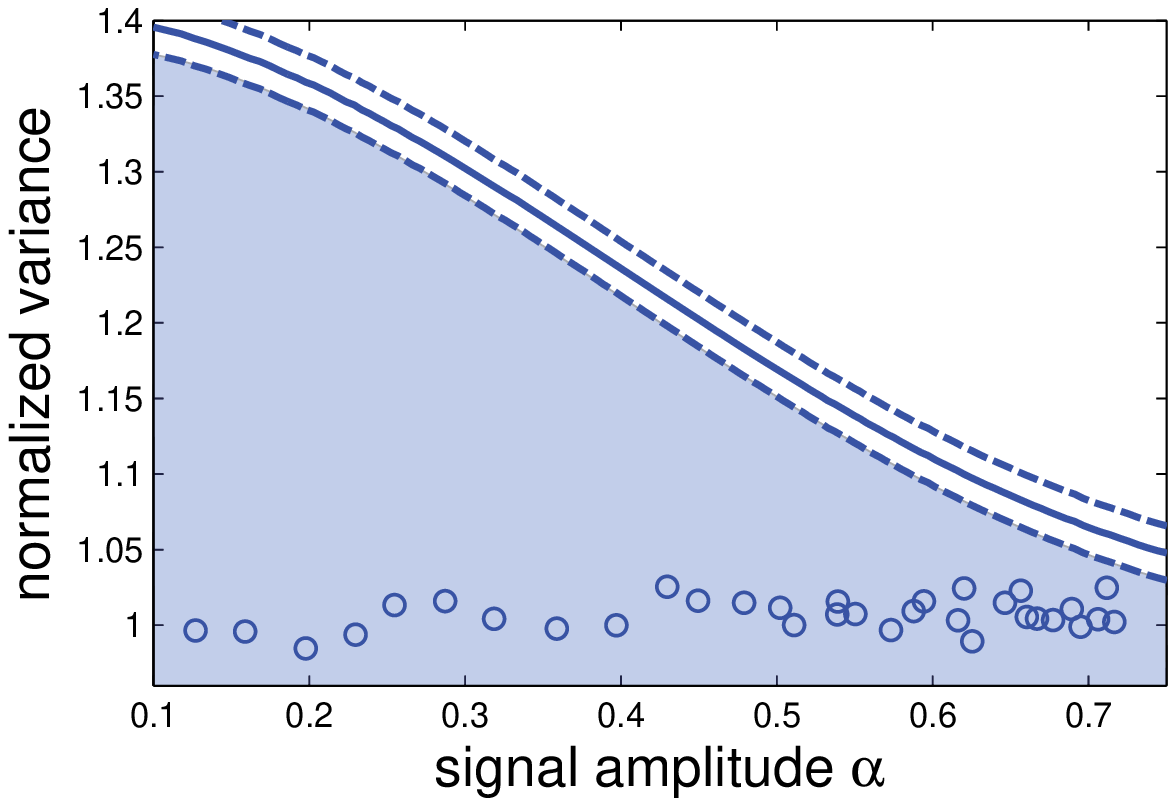}}\\[-10.3cm]
\hspace{0.5cm}(a)\hspace{6cm}(b)\\[5cm]
(c)\hspace{6.1cm}(d)\\[4.4cm]
\end{tabular}
\caption{\label{fig:results2m}  Alice and Bob are connected back to back. (a) Combined Q-function generated from 5 million pulses with amplitudes $\alpha_S=0.5$  (b) Q-functions for each signal state. (c) Excess noise estimated from Q-function for varying signal amplitudes. (d) Average excess noise (circles) compared to the bounds given by the entanglement criterion (solid line), shown with the 3-sigma confidentiality interval (dashed lines).}
\end{figure}

Reformulating the measurement in the Stokes representation (see Eqns.~(\ref{Stokesoperators1})) is of great importance for the security analysis of the system.
We find that Eqns.~(\ref{HD1}) and (\ref{HD2}) have the form of a simultaneous Stokes measurement of the $\hat S_\phi$- and the $\hat S_{\phi+\pi/2}$-operator, respectively, where $\hat S_\theta=\hat S_2\cos\theta+\hat S_3\sin\theta$. The Stokes operator $\hat S_1$ is estimated by the monitor diode, as shown in Eqn.~(\ref{S1est}).

To verify the effective entanglement in our experiment, we compare the measured excess noise to the theoretical upper noise bounds (see Fig.~\ref{fig:theory}). The excess noise estimation demands a calibration step. The standard procedure to calibrate the shotnoise level is to probe vacuum. Therefore, we physically block the signal arm of the detection stage and monitor the noise power of the Stokes measurements, as well as the signal of the LO monitor diode. The shotnoise estimation thus corresponds to a measurement of the electronic gains in the detection system using a well known and low noise light beam\footnote{If the LO has traveled through the eavesdropper's regime, it is mandatory to verify that the beam exhibits small excess noise compared to the detector's common mode rejection ratio, e.g. with an attenuation measurement~\cite{bachor_guide_2004}.}. In standard homodyne detection, the variance of the shot noise will then scale with the power of the LO in following measurements. However, a new aspect is introduced in Eqn.~(\ref{S1est}): If there is significant noise power in the $\hat S_2$- or $\hat S_3$-component of the signal states, the bound on $\braket{\hat{S}_1}$ will deteriorate, and with it the ability to detect quantum correlations.  We estimate the contribution of the last term in Eqn.~(\ref{S1est}). We consider the worst case, i.e. the maximal reduction of the expectation value in each measurement run for a single amplitude (5 million signal pulses).
\begin{equation}
\frac{\braket{\hat{S}_{2}^2}}{\braket{\hat{n}_{LO}}}+\frac{\braket{\hat{S}_{3}^2}}{\braket{\hat{n}_{LO}}}\leq \frac{\gamma_\mathrm{LO}}{\gamma_{S_2}^2}\max\left(\frac{\braket{\hat{u}_{S_2}^2}}{\braket{\hat{u}_{LO}}}\right)+\frac{\gamma_\mathrm{LO}}{\gamma_{S_3}^2}\max\left(\frac{\braket{\hat{u}_{S_3}^2}}{\braket{\hat{u}_{LO}}}\right),
\label{snlreduction}
\end{equation}
where $\gamma_{S_i}$, $\hat{u}_{S_i}$, $\gamma_\mathrm{LO}$ and $\hat{u}_{\mathrm{LO}}$ are gains and output voltages of the detectors for the $\hat S_i$-components ($i=2,3$) and the LO power, respectively. The calibration measurement is used to measure the gain ratio. For an approximately shot noise limited beam we find 

\begin{equation}
\frac{\gamma_\mathrm{LO}}{\gamma_{S_i}^2}\frac{\Delta^2\hat{u}_{S_i,\mathrm{snl}}}{\braket{\hat{u}_{LO,\mathrm{snl}}}}= \frac{\Delta^2\hat{S}_{i,\mathrm{snl}}}{\braket{\hat{n}_{LO,\mathrm{snl}}}}=1.
\label{snlreduction2}
\end{equation}

The reduction of the normalized expectation value is fluctuating during the measurements. However, using Eqns.~(\ref{snlreduction}) and (\ref{snlreduction2}), we calculated that it never surpasses $10^{-3}$. This factor is considered by shifting the bound for the entanglement verification. Another uncertainty in the level of shotnoise is the estimation error of the LO power. The LO power is measured for every signal pulse separately and therefore has a rather large standard deviation of $6\cdot10^{-3}$. We consider the deviation by plotting the three sigma confidence interval around the upper noise bound.

\begin{figure}
\begin{tabular}{l}
\centerline{\includegraphics[width=7.0cm]{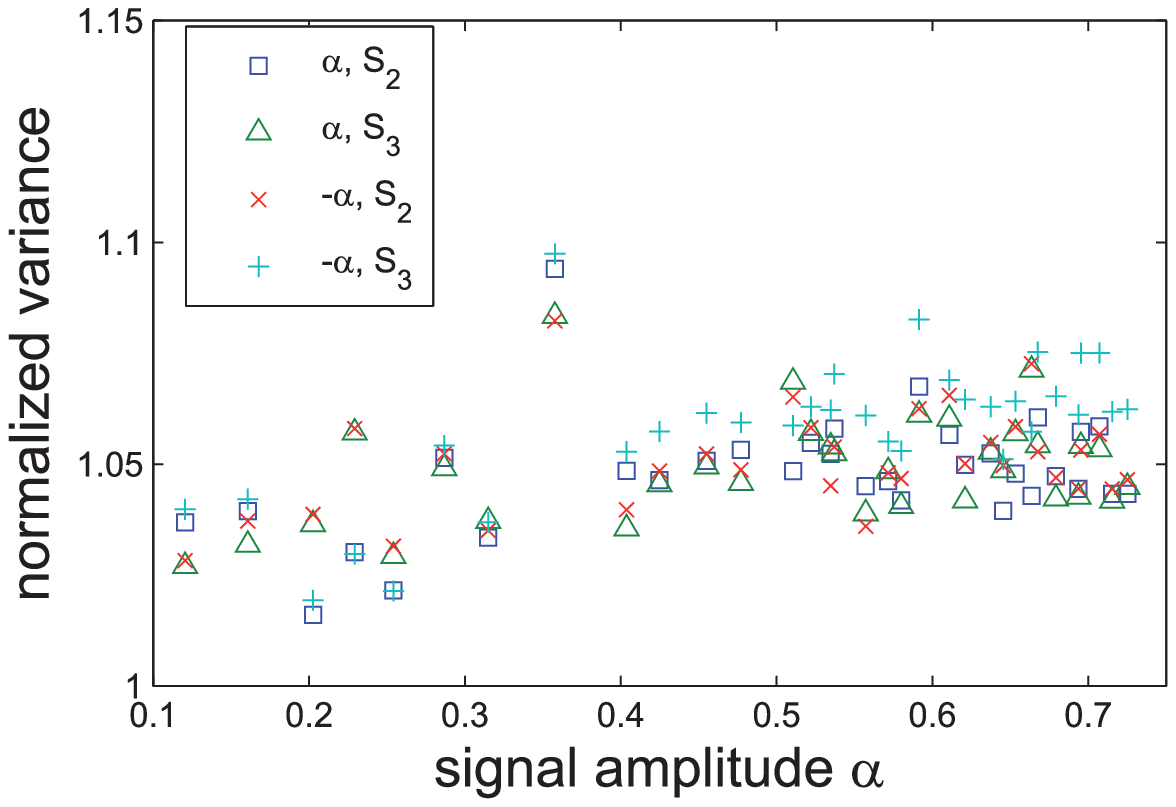}\hspace{-0.4cm}\includegraphics[width=7.0cm]{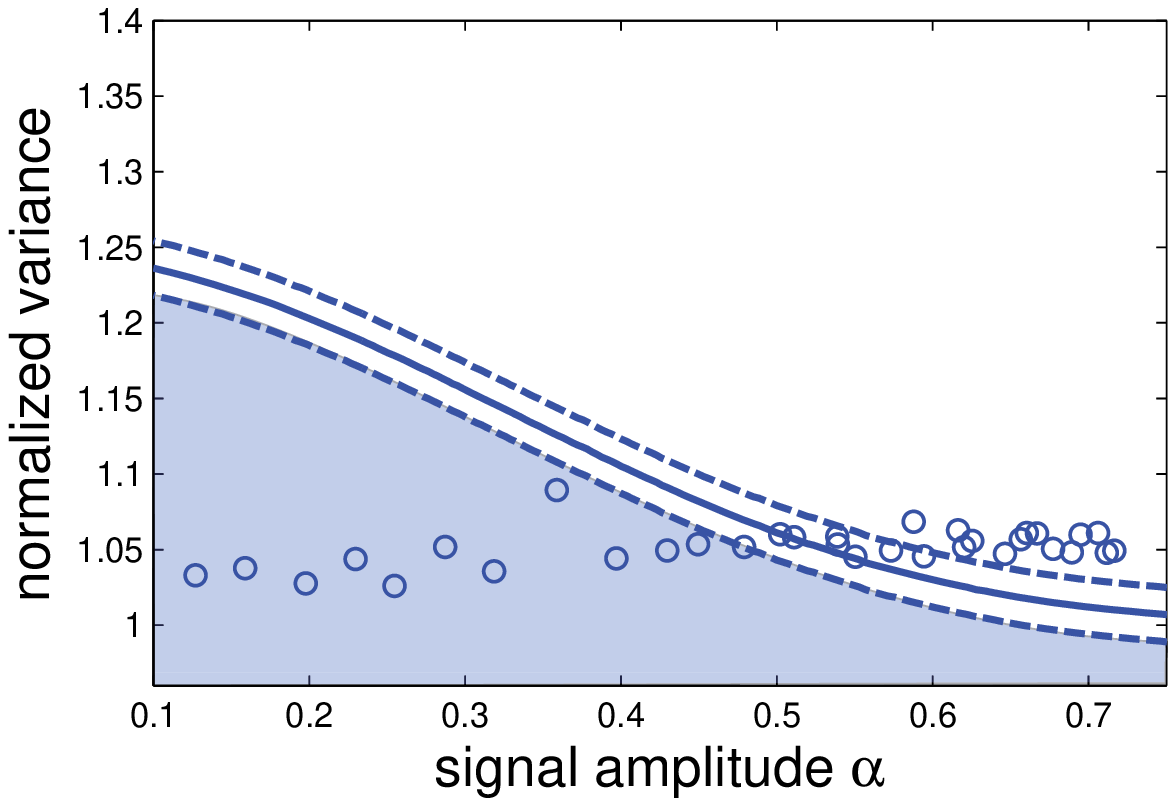}}\\[-4.8cm]
(a)\hspace{6.1cm}(b)\\[4.3cm]
\end{tabular}
\caption{\label{fig:results2km}  Alice and Bob are connected with a 2\,km optical fiber. (a) Excess noise estimated from Q-function measurement for varying signal amplitudes. (b) Average excess noise (circles) compared to the bounds given by the entanglement criterion (solid line), shown with the 3-sigma confidentiality interval (dashed lines).}
\end{figure}

In our first measurement, we connected both modules back to back. The effective detection efficiency is therefore merely the $70\%$ efficiency of the receiver module. The Q-functions are measured for the two signal states combined or separately as shown in Fig.~\ref{fig:results2m}(a) and (b), respectively. From the Q-functions, we estimate the mean value and the variance for both Stokes opertors of both signal states as shown in Fig.~\ref{fig:results2m}(c). We find that the variances are approximately equal. Their average variance is therefore compared to the upper bound of the variance given by the Expectation Value Matrix method. The comparison of upper bound and average variance is shown in~\ref{fig:results2m}(d). We find, that for all tested signal amplitudes the measured noise is below  the 3-sigma confidentiality interval. Therefore the effective entanglement is verified.

We then substitute the 2\,m quantum channel with a 2\,km standard optical fiber. The effective detection efficiency was $44.8\%$. The transmission of the channel decreases to $64\%$ also due to additional splices. In an optimized setting, this attenuation corresponds to at least 10\,km of fiber. Again, we estimate the mean values and the variance for both Stokes opertors of both signal states as shown in Fig.~\ref{fig:results2km}(a). Finally, we compare the average values to the upper bound of the variance given by the Expectation Value Matrix method (Fig.~\ref{fig:results2km}(b)). We find that for signal amplitudes below 0.45, effective entanglement is verified in our system.

\section{Conclusion}
In conclusion, we present a fiber-based continuous-variable quantum key distribution system. We demonstrate a receiver module, which simultaneously measures conjugate Stokes operators of light. This is the first simultaneous detection of conjugate Stokes operator of a quantum signal after a fiber channel. From the measured Q-function of the transmitted signal, we estimate the attenuation and the excess noise caused by the channel. For the measured amount of excess noise, the theory has not progressed far enough to generate an unconditionally secure and secret key. Nevertheless, we successfully witness effective entanglement with a channel length of up to $2 \unit{km}$ considering both parts of the quantum signal, the signal and the LO mode.

\section{Acknowledgement}
The authors would like to thank Georgy Onishchukov for fruitful discussions and complementary hardware.


\begin{thebibliography}{10}
\newcommand{\enquote}[1]{``#1''}

\bibitem{kerckhoffs_la_1883}
A.~Kerckhoffs, \emph{La cryptographie militaire} (Journal des sciences
  militaires, Vol. {IX,} pp. 5-38, 1883).

\bibitem{shannon_communication_1949}
C.~Shannon, \enquote{Communication theory of secrecy systems,} The Bell system
  technical journal \textbf{28}, 656--715 (1949).

\bibitem{vernam_cipher_1926}
G.~Vernam, \enquote{Cipher printing telegraph systems for secret wire and radio
  telegraphic communications,} J. Amer. Inst. Elect. Eng. p. 109 (1926).

\bibitem{bennett_quantum_1984}
C.~Bennett and G.~Brassard, \enquote{Quantum cryptography: Public key
  distribution and coin tossing,} Proceedings of {IEEE} International
  Conference on Computers Systems and Signal Processing, Bangarore India pp.
  175--179 (1984).

\bibitem{bennett_experimental_1992}
C.~Bennett, F.~Bessette, G.~Brassard, L.~Salvail, and J.~Smolin,
  \enquote{Experimental quantum cryptography,} Journal of Cryptology
  \textbf{5}, 3--28 (1992).

\bibitem{ekert_practical_1992}
A.~Ekert, J.~Rarity, P.~Tapster, and G.~Palma, \enquote{Practical quantum
  cryptography based on 2-photon interferometry,} Phys. Rev. Lett.
  \textbf{69}, 1293--1295 (1992).

\bibitem{muller_experimental_1993}
A.~Muller, J.~Breguet, and N.~Gisin, \enquote{Experimental demonstration of
  quantum cryptography using polarized photons in optical-fiber over more than
  1 km,} Europhys. Lett.  \textbf{23}, 383--388 (1993).

\bibitem{stucki_quantum_2002}
D.~Stucki, N.~Gisin, O.~Guinnard, G.~Ribordy, and H.~Zbinden, \enquote{Quantum
  key distribution over 67 km with a plug\&play system,} New J. Phys.
  \textbf{4}, 41 (2002).

\bibitem{rosenberg_long-distance_2007}
D.~Rosenberg, J.~W. Harrington, P.~R. Rice, P.~A. Hiskett, C.~G. Peterson,
  R.~J. Hughes, A.~E. Lita, S.~W. Nam, and J.~E. Nordholt,
  \enquote{Long-distance decoy-state quantum key distribution in optical
  fiber,} Phys. Rev. Lett. \textbf{98}, 10503 (2007).

\bibitem{ursin_entanglement-based_2007}
R.~Ursin, F.~Tiefenbacher, T.~{Schmitt-Manderbach}, H.~Weier, T.~Scheidl,
  M.~Lindenthal, B.~Blauensteiner, T.~Jennewein, J.~Perdigues, P.~Trojek,
  B.~\"Omer, M.~F\"urst, M.~Meyenburg, J.~Rarity, Z.~Sodnik, C.~Barbieri,
  H.~Weinfurter, and A.~Zeilinger, \enquote{Entanglement-based quantum
  communication over 144km,} Nat. Phys.  \textbf{3}, 481--486 (2007).

\bibitem{schmitt-manderbach_experimental_2007}
T.~{Schmitt-Manderbach}, H.~Weier, M.~F\"urst, R.~Ursin, F.~Tiefenbacher,
  T.~Scheidl, J.~Perdigues, Z.~Sodnik, C.~Kurtsiefer, J.~G. Rarity,
  A.~Zeilinger, and H.~Weinfurter, \enquote{Experimental demonstration of
  free-space decoy-state quantum key distribution over 144 km,} Phys. Rev. Lett. \textbf{98}, 010504 (2007).

\bibitem{ralph_continuous_1999}
T.~C. Ralph, \enquote{Continuous variable quantum cryptography,} Phys.
  Rev. A \textbf{61}, 010303 (1999).

\bibitem{silberhorn_continuous_2002}
C.~Silberhorn, T.~C. Ralph, N.~L\"utkenhaus, and G.~Leuchs, \enquote{Continuous
  variable quantum cryptography: Beating the 3 {dB} loss limit,} Phys.
  Rev. Lett. \textbf{89}, 167901 (2002).

\bibitem{grosshans_quantum_2003}
F.~Grosshans, G.~V. Assche, J.~Wenger, R.~Brouri, N.~J. Cerf, and P.~Grangier,
  \enquote{Quantum key distribution using gaussian-modulated coherent states,}
  Nature \textbf{421}, 238--241 (2003).

\bibitem{lodewyck_quantum_2007}
J.~Lodewyck, M.~Bloch, R.~{Garcia-Patron}, S.~Fossier, E.~Karpov, E.~Diamanti,
  T.~Debuisschert, N.~J. Cerf, R.~{Tualle-Brouri}, S.~W. {McLaughlin}, and
  P.~Grangier, \enquote{Quantum key distribution over 25 km with an all-fiber
  continuous-variable system,} Phys. Rev. A \textbf{76}, 042305--10
  (2007).

\bibitem{qi_experimental_2007}
B.~Qi, L.~Huang, L.~Qian, and H.~Lo, \enquote{Experimental study on the
  gaussian-modulated coherent-state quantum key distribution over standard
  telecommunication fibers,} Phys. Rev. A \textbf{76}, 052323--9 (2007).

\bibitem{lorenz_continuous-variable_2004}
S.~Lorenz, N.~Korolkova, and G.~Leuchs, \enquote{Continuous-variable quantum
  key distribution using polarization encoding and post selection,} Appl. Phys. B \textbf{79}, 273--277 (2004).

\bibitem{lance_no-switching_2005}
A.~M. Lance, T.~Symul, V.~Sharma, C.~Weedbrook, T.~C. Ralph, and P.~K. Lam,
  \enquote{No-switching quantum key distribution using broadband modulated
  coherent light,} Phys. Rev. Lett. \textbf{95}, 180503--4 (2005).

\bibitem{gisin_trojan-horse_2006}
N.~Gisin, S.~Fasel, B.~Kraus, H.~Zbinden, and G.~Ribordy, \enquote{Trojan-horse
  attacks on quantum-key-distribution systems,} Phys. Rev. A \textbf{73},
  022320--6 (2006).

\bibitem{pirandola_continuous-variable_2008}
S.~Pirandola, S.~Mancini, S.~Lloyd, and S.~L. Braunstein,
  \enquote{Continuous-variable quantum cryptography using two-way quantum
  communication,} Nat. Phys. \textbf{4}, 726--730 (2008).

\bibitem{weedbrook_coherent-state_2006}
C.~Weedbrook, A.~M. Lance, W.~P. Bowen, T.~Symul, T.~C. Ralph, and P.~K. Lam,
  \enquote{Coherent-state quantum key distribution without random basis
  switching,} Phys. Rev. A \textbf{73}, 022316--9 (2006).

\bibitem{lodewyck_tight_2007}
J.~Lodewyck and P.~Grangier, \enquote{Tight bound on the coherent-state quantum
  key distribution with heterodyne detection,} Phys. Rev. A \textbf{76},
  022332--8 (2007).

\bibitem{elser_feasibility_2009}
D.~Elser, T.~Bartley, B.~Heim, C.~Wittmann, D.~Sych, and G.~Leuchs,
  \enquote{Feasibility of free space quantum key distribution with coherent
  polarization states,} New J. Phys. \textbf{11}, 045014 (2009).

\bibitem{elser_guided_2007}
D.~Elser, C.~Wittmann, U.~L. Andersen, O.~Gl\"ockl, S.~Lorenz, C.~Marquardt, and
  G.~Leuchs, \enquote{Guided acoustic wave brillouin scattering in photonic
  crystal fibers,} J. Phys. Conf. Ser. \textbf{92}, 012108
  (2007).

\bibitem{korolkova_polarization_2002}
N.~Korolkova, G.~Leuchs, R.~Loudon, T.~C. Ralph, and C.~Silberhorn,
  \enquote{Polarization squeezing and continuous-variable polarization
  entanglement,} Phys. Rev. A \textbf{65}, 052306 (2002).

\bibitem{hseler_testing_2008}
H.~H\"aseler, T.~Moroder, and N.~L\"utkenhaus, \enquote{Testing quantum devices:
  Practical entanglement verification in bipartite optical systems,} Phys.
  Rev. A \textbf{77}, 032303--11 (2008).

\bibitem{dorrer_linear_2003}
C.~Dorrer, D.~Kilper, H.~Stuart, G.~Raybon, and M.~Raymer, \enquote{Linear
  optical sampling,} {IEEE} Photonics Technol. Lett. \textbf{15},
  1746--1748 (2003).

\bibitem{bennett_quantum_1992-1}
C.~H. Bennett, \enquote{Quantum cryptography using any two nonorthogonal
  states,} Phys. Rev. Lett. \textbf{68}, 3121 (1992).

\bibitem{zhao_asymptotic_2009}
Y.~Zhao, M.~Heid, J.~Rigas, and N.~L\"utkenhaus, \enquote{Asymptotic security of
  binary modulated continuous-variable quantum key distribution under
  collective attacks,} Phys. Rev. A \textbf{79}, 012307--14 (2009).

\bibitem{curty_entanglement_2004}
M.~Curty, M.~Lewenstein, and N.~L\"utkenhaus, \enquote{Entanglement as a
  precondition for secure quantum key distribution,} Phys. Rev. Lett.
  \textbf{92}, 217903 (2004).

\bibitem{bennett_quantum_1992}
C.~H. Bennett, G.~Brassard, and N.~D. Mermin, \enquote{Quantum cryptography
  without bell's theorem,} Phys. Rev. Lett. \textbf{68}, 557 (1992).

\bibitem{rigas_entanglement_2006}
J.~Rigas, O.~G\"uhne, and N.~L\"utkenhaus, \enquote{Entanglement verification for
  quantum-key-distribution systems with an underlying bipartite qubit-mode
  structure,} Phys. Rev. A \textbf{73}, 012341--6 (2006).

\bibitem{leonhardt_measuringquantum_1997}
U.~Leonhardt, \emph{Measuring the Quantum State of Light} (Cambridge University
  Press, 1997).

\bibitem{legr_implementation_2006}
M.~Legr\' e, H.~Zbinden, and N.~Gisin, \enquote{Implementation of continuous
  variable quantum cryptography in optical fibres using a go-\&-return
  configuration,} Quantum Inf. Comput. \textbf{6}, 326--335
  (2006).

\bibitem{nelder_simplex_1965}
J.~A. Nelder and R.~Mead, \enquote{A simplex method for function minimization,}
  The Computer Journal \textbf{7}, 308--313 (1965).

\bibitem{shapiro_phase_1984}
J.~Shapiro and S.~Wagner, \enquote{Phase and amplitude uncertainties in
  heterodyne detection,} {IEEE}  J. Quantum Electron. \textbf{20},
  803--813 (1984).

\bibitem{stenholm_simultaneous_1992}
S.~Stenholm, \enquote{Simultaneous measurement of conjugate variables,} Ann. Phys. \textbf{218}, 233--254 (1992).

\bibitem{leonhardt_realistic_1993}
U.~Leonhardt and H.~Paul, \enquote{Realistic optical homodyne measurements and
  quasi-probability distributions,} Phys. Rev. A \textbf{48}, 4598--4604
  (1993).

\bibitem{hansen_ultrasensitive_2001}
H.~Hansen, T.~Aichele, C.~Hettich, P.~Lodahl, A.~Lvovsky, J.~Mlynek, and
  S.~Schiller, \enquote{Ultrasensitive pulsed, balanced homodyne detector:
  application to time-domain quantum measurements,} Opt. Lett. \textbf{26},
  1714--1716 (2001).

\bibitem{leonhardt_canonical_1995}
U.~Leonhardt, J.~A. Vaccaro, B.~B\"ohmer, and H.~Paul, \enquote{Canonical and
  measured phase distributions,} Phys. Rev. A \textbf{51}, 84 (1995).

\bibitem{bachor_guide_2004}
H.~Bachor and T.~Ralph, \emph{A guide to experiments in quantum optics}
  ({Wiley-VCH} Verlag, 2004).

\end{thebibliography}
\end{document}